# Uncertainty-aware phase fraction prediction and active-learning-guided out-of-domain discovery of refractory multi-principal element alloys

Ali K. Shargh[1*], Christopher D. Stiles[1,2], Jaafar A. El-Awady[1†]

[1] Department of Mechanical Engineering, Johns Hopkins University, Baltimore, Maryland 21218, United States

[2] Research and Exploratory Development Department, Johns Hopkins Applied Physics Laboratory, Laurel, Maryland 20723, United States

## Abstract

Refractory multi-principal element alloys (RMPEAs) represent a novel class of alloys characterized by an extensive compositional design space and the potential for exceptional mechanical performance under extreme conditions. While accurate phase stability prediction is essential for their robust design, existing machine learning approaches rely on deterministic mappings from composition-derived features to phase labels, neglecting the uncertainty inherent in such predictions. In this study, we present a deep learning framework based on Mixture Density Networks (MDNs) to predict phase fractions in RMPEAs and quantify the associated aleatoric uncertainty across a wide temperature range. By training separate models for up to six constituent phases of RMPEAs using CALPHAD derived data, our approach achieves high predictive accuracy while capturing the probabilistic nature of phase formation. To address epistemic uncertainty arising from incomplete knowledge of the most informative features, we perform a perturbation-based feature importance analysis and identify a minimally sufficient input set that maintains both predictive performance and uncertainty calibration. Finally, we propose an uncertainty-based active learning strategy to discover novel RMPEAs with the target phase incorporating previously unseen elements, while investigating the exploration–exploitation trade-off in model-guided discovery. Our uncertainty-aware framework has the potential to accelerate and improve the reliability of discovering novel high-performance alloys and is broadly applicable.

---

[*] Contact author: ashargh1@jhu.edu (A. K. Shargh)
[†] Contact author: jelawady@jhu.edu (J. A. El-Awady)

# I. Introduction

Refractory multi-principal-element alloys (RMPEAs) consist of multiple principal elements in non-dilute concentrations [1–3] that offer a vast compositional design space and have the potential to exhibit remarkable mechanical performance in extreme environments [4]. Navigating this design space is hindered by sparse experimental data and the complicated relationships between composition and properties.

To overcome these hurdles, data-driven and machine-learning (ML) surrogate models have been developed for accelerating materials discovery and design, including in RMPEAs through property prediction [5–10]. Among various predicted properties, phase stability is especially critical for RMPEAs, as it directly influences their mechanical behavior. For example, the face-centered cubic (FCC) structure is known to promote enhanced ductility [11], the body-centered cubic (BCC) structure improves strength [12], intermetallic Laves phases increase wear resistance [13], and sigma (σ) phases reduce corrosion resistance [14].

In our earlier studies [15,16], we developed deep-learning (DL) models as a surrogate for CALculation of PHAse Diagrams (CALPHAD) that achieve high accuracy in classifying the constituent RMPEAs phases and predicting their fractions across a range of temperatures using up to 35 physicochemical input features. Our models accelerated the phase prediction of RMPEAs by two orders of magnitude compared to CALPHAD, making large-scale design space exploration computationally feasible, which would be impractical using CALPHAD alone.

Despite advances in ML-assisted phase prediction of RMPEAs, a key limit remains largely unaddressed. Previous ML models [17–23], including our own [15,16], are designed to learn a deterministic mapping from input features to output phase labels. However, different compositions can produce nearly identical feature values, meaning that distinct compositions may be mapped to the same point in feature space while exhibiting different phase constitutions.

This inherent ambiguity gives rise to aleatoric uncertainty, which reflects the complex variability in phase stability predictions due to overlapping or non-unique representations in the feature space. This critical source of uncertainty has not been addressed in prior phase prediction studies. In addition, existing studies on phase prediction often employ different combinations of input features, leading to inconsistencies in feature selection across literature. This variation reflects the fact that our understanding of the most informative features is still evolving. As new descriptors are discovered and incorporated, models trained on earlier or incomplete feature sets may suffer from uncertainty due to missing or suboptimal input information. This type of uncertainty, referred to as epistemic uncertainty, stems from incomplete knowledge about the optimal set of input features and must be addressed to improve model robustness. The current study thus seeks to address this gap by using DL models to quantify aleatoric and epistemic uncertainties in RMPEA phase prediction, which are respectively associated with data variability and incomplete knowledge of the input space. Specifically, we employ a Mixture Density Network (MDN) [24] to predict phase fractions and their associated aleatoric uncertainty over a temperature range of 850–1892 K for 3-, 4-, and 5-element alloys within the Ti, Fe, Al, V, Ni, Nb, Zr, Mn, and Co elemental space. The focus on Ti, Fe, Al, V, Ni, Nb, Zr, Mn, and Co in this study reflects their growing importance as key constituents of RMPEAs for high-temperature applications [25,26]. Once trained, the framework is used to perform a systematic feature importance analysis that identifies key input descriptors. We then quantify how reducing the number of input features affects both predictive accuracy and epistemic uncertainty, providing insights into the model's sensitivity to feature selection. It is noteworthy that uncertainty quantification remains an underexplored aspect of RMPEA research [27], with only a few studies addressing it across different contexts [28–31]. For example, Zhang et al. [30] proposed an ML framework grounded in Bayesian approaches to predict the configurational energy of RMPEAs. Their uncertainty quantification of effective pair interactions revealed significant variation and frustration within the nearest-neighbor shell, which reduced as more data were incorporated, indicating that bond-level uncertainties stabilize with increased training data.

Another objective of this study is to develop an uncertainty-based active learning approach to discover novel RMPEAs with the target phase, incorporating elements previously unseen by the model. Conventional DL models, such as multilayer perceptron (MLP) models, often produce overconfident predictions when extrapolating to unfamiliar regions of the design space. This overconfidence makes it challenging to distinguish between meaningful extrapolations and unreliable guesses. In contrast, MDN models provide explicit uncertainty quantification, allowing the model to recognize low-reliability regions and guide the exploration of previously unseen areas of the design space with higher confidence, which is particularly advantageous in the discovery of novel materials [32,33]. As such, this direction is particularly relevant in ML-assisted high-throughput experimental settings, where researchers must explore new material systems that differ substantially in composition, structure, or properties from those in the training dataset to discover novel materials. These differences can lead to shifts in data distribution, making the target materials effectively out-of-distribution (OOD) for the ML models. Given that most ML models depend on representative training data, achieving robust OOD generalization remains a significant challenge in materials discovery. In recent years, several studies have aimed to highlight and address OOD prediction challenges in materials science via active learning methods [34–40]. Active learning methods iteratively sample from the design space and incorporate newly acquired data to refine model predictions and guide future sampling. For example, Borg et al. [35] proposed a framework based on active learning to accelerate the discovery of materials with target band gap values. Their study demonstrated that the framework's performance is sensitive to factors such as the target range of band gap values, the number of iterations, the number of desired discoveries, and the type of acquisition functions.

This paper is organized to systematically address the objectives described above. Section II details the processes of database construction, data labeling, and the design of the DL framework used in this study. Section III presents a comprehensive evaluation of the framework, focusing on three main aspects: (1) assessment of model accuracy and aleatoric uncertainty quantification; (2)

analysis of feature importance and corresponding epistemic uncertainty; and (3) out-of-distribution (OOD) discovery of novel RMPEAs using an uncertainty-based active learning approach. Finally, the Conclusion section summarizes the key findings and discusses their broader implications for accelerated and reliable RMPEA design.

## II. Methods

### A. Dataset generation

Before training our deep-learning model, we need to construct a representative dataset from the design space and label it with input features and the corresponding target RMPEA phases. To that end, we use a random sampling approach to generate 70,000 unique compositions from the design space of our elements of interest (Ti, Fe, Al, V, Ni, Nb, Zr, Mn, Co). Each composition contains 3 to 5 elements, with elemental fractions spanning 5% to 35% in 1% intervals.

We then compute the target RMPEA phases for each composition at seven temperatures ranging from 850 K to 0.8 $T_m$ using the Thermo-Calc Python interface and its thermodynamic database TCHEA6 [41]. The melting temperature $T_m$ for each alloy is estimated based on the mean mixture rule. These seven uniformly spaced temperatures between 850 K and 0.8 $T_m$ are unique to each RMPEA, as $T_m$ varies from one sample to another. It should be noted that Thermo-Calc fails to find a solution for 5,036 of the 490,000 RMPEAs. These compositions are subsequently removed from the final dataset. Importantly, our focus is on the FCC, BCC, B2, HCP, Laves (C14, C15, and C36), Heusler, Sigma, and Liquid phases. Due to the limited availability of HCP phase data within the selected element space, we do not include this phase in our analysis. In addition, BCC and B2 are grouped. As discussed in our earlier study [16], combining the BCC and B2 phases into a single regression target improves the predictive performance of DL models.

When distinct phases exhibit identical crystal structures but differ in composition, such as those resulting from spinodal decomposition, they are grouped under a standard structural label. For instance, a composition containing both Laves#1 and Laves#2 is categorized simply as Laves. Because the phase distribution is heavily imbalanced across all phases, as shown in Figure 1, we apply a custom oversampling strategy to improve training effectiveness in this study. In our oversampling procedure, the phase-fraction distribution of the training dataset is discretized into bins of width 0.1. Each bin is then oversampled to contain at least half as many samples as the most populated original bin. The resulting phase-fraction distribution is shown in Supplementary Figure S1. Importantly, most existing strategies for addressing data imbalance have been developed for classification problems, whereas analogous approaches for regression tasks remain relatively limited. For comparison, we have also tested other oversampling schemes proposed by Branco et al. [42] including Gaussian noise injection, SMOTER, WERCS, balanced random oversampling, and extreme random oversampling for the FCC phase (see Supplementary Table I in Supplementary Section I). However, these alternative methods did not yield a noticeable improvement in model performance.

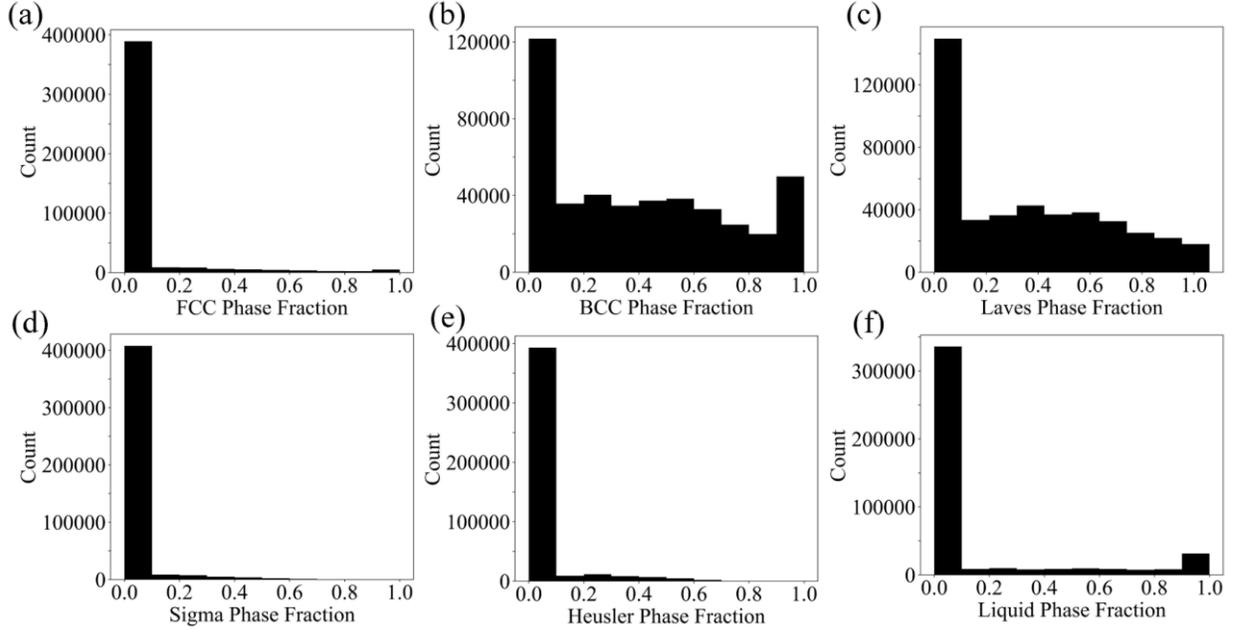

**Figure 1:** Distributions of RMPEA phase fractions in the original dataset excluding the testing dataset.

Next, we label the samples with 51 features, selected based on a comprehensive set we developed in our earlier studies [15,16] through an extensive review of the literature on phase prediction in RMPEAs. These 51 input features include: mixing entropy $\Delta S_{\text{mix}}$, mixing enthalpy $\Delta H_{\text{mix}}$, $\Omega$, $\eta$, $k_1^{\text{cr}}$, atomic size difference $\delta$, the valence electron concentration VEC, electronegativity $\chi$, the phase formation parameters (PFP) for phases $PFP_{\text{FCC}}$, $PFP_{\text{BCC}}$, $PFP_{\text{HCP}}$, $PFP_{B_2}$, $PFP_{\text{Laves}}$, $PFP_{\text{Sigma}}$, and the phase separation parameter PSP [43], bulk modulus K, melting temperature $T_{\text{m}}$, the standard deviations of mixing enthalpy $\sigma_{\Delta H_{\text{mix}}}$, bulk modulus $\sigma_K$, melting temperature $\sigma_{T_{\text{m}}}$, and valence electron concentration $\sigma_{VEC}$, standard deviation of electronegativity $\Delta\chi$, atomic number, group (i.e., the vertical columns numbered 1 through 18 in the periodic table), families (i.e. Alkali metals, Alkaline-earth metals, Rate-earth elements, Transition metals, basic metals, semi-metals, Nonmetals, Halogens, Noble gases), $\Phi$ which quantifies the change in Gibbs free energy for the formation of a solid solution phase, normalized by the lowest possible Gibbs free energy from binary systems, and is calculated using the publicly available Alloy Search and Predict (ASAP) code [44], quantum number L, miracle radius, covalent radius, Zunger radius, ionic radius, crystal radius, Mulliken-Badger (MB) electronegativity, Gordy electronegativity, Allred-Rockow electronegativity, polarizability, boiling point, density, specific heat, thermal conductivity, the temperature $T$, as well as $\frac{E_2}{E_0}$, atomic weight, period, Mendeleev number, Mulliken electronegativity, first ionization potential, heat of fusion, heat of vaporization, heat atomization, and cohesive energy. The equations used to derive the features are provided in Table .

**Table I:** Summary of equations used to calculate selected input features.

| Parameter | Equation | Descriptions |
|---|---|---|
| $\Delta S_{\text{mix}}$ | $-R \sum_{i=1}^{N} c_i \ln(c_i)$ | $R$ is the gas constant which is equal to 8.314 J/(mol.T), and $c_i$ is the concentration of element $i$ in atomic fraction. |
| $\Delta H_{\text{mix}}$ | $\sum_{i=1,\ i \neq j}^{N} 4\Delta H_{ij}^{\text{mix}} c_i c_j$ | $\Delta H_{ij}^{\text{mix}}$ are calculated from available tables that were obtained by Miedema's model [45]. |
| $\Omega$ | $\dfrac{T_m \Delta S_{\text{mix}}}{|\Delta H_{\text{mix}}|}$ | $T_m$ is calculated from $\sum_{i=1}^{N} c_i T_i^m$ wherein $T_i^m$ is the melting temperature of element $i$ |

| | | |
|---|---|---|
| $\eta$ | $\dfrac{-T_{\text{ann}}\Delta S_{\text{mix}}}{|\Delta H_{\text{f}}|}$ | $T_{\text{ann}}$ is estimated as $0.8T_{\text{m}}$, and $\Delta H_{\text{f}}$ is the most negative binary mixing enthalpy for forming intermetallic phase (i.e. $H_{ij}^{\text{IM}}$) that are reported in [46] |
| $k_1^{\text{cr}}$ | $\dfrac{\left(1 - \dfrac{0.4T_{\text{m}}\Delta S_{\text{mix}}}{|\Delta H_{\text{mix}}|}\right)}{\dfrac{\Delta H_{\text{IM}}}{\Delta H_{\text{mix}}}}$ | $\Delta H_{\text{IM}}$ is mixing enthalpy for forming intermetallic phase |
| $\delta$ | $100 \times \sqrt{\sum_{i=1}^{N} c_i [1 - \dfrac{r_i}{\sum_{j=1}^{N} c_j r_j}]^2}$ | $r_i$ is the atomic radius of element $i$. |
| $\Delta\chi$ | $\sqrt{\sum_{i=1}^{N} c_i \left[\chi_i - \sum_{j=1}^{N} c_j \chi_j\right]^2}$ | $\chi_i$ is Electronegativity of element $i$ |
| $\dfrac{E_2}{E_0}$ | $\sum_{j \geq i}^{N} \dfrac{c_i c_j |r_i + r_j - 2\bar{r}|^2}{(2\bar{r})^2}$ | |
| VEC, $\chi$, K, $T_{\text{m}}$, atomic number, group, families, quantum number L, miracle radius, covalent radius, Zunger radius, ionic radius, crystal radius, Mulliken-Badger (MB) | $x_{\text{avg}} = \sum_{i=1}^{N} c_i x_i$ | $c_i$ is the concentration of element $i$ in atomic fraction and $x_i$ is the parameter value for element $i$. The values of $x_i$ are publicly available [47,48] |

| electronegativity, Gordy electronegativity, Allred-Rockow electronegativity, polarizability, boiling point, density, specific heat, thermal conductivity, atomic weight, period, Mendeleev number, Mulliken electronegativity, first ionization potential, heat of fusion, heat of vaporization, heat atomization, and cohesive energy | | |
|---|---|---|

The initial set of 51 features is then refined to 41 distinct features through a two-step data engineering strategy, with the 51 features initially presented in an order that places the 10 excluded features at the end to avoid repeating the full list, as follows: 1) To address redundancy and mitigate multicollinearity, we perform pairwise correlation analysis and remove one feature from each pair with a strong linear relationship (|Pearson correlation coefficient| > 0.9). (2) To improve feature distribution uniformity, all features are scaled to the [0, 1] range using min-max normalization based on the observed dataset range. We then evaluate the distribution of each feature by constructing histograms with a bin width of 0.1 and removing data points that fall into bins containing fewer than 20 samples. This filtering step removes 899 compositions. Combined with the 5,036 compositions removed earlier due to failed Thermo-Calc solutions (out of the original 490,000), the final dataset consists of 484,065 compositions.

## B. Deep learning model with uncertainty

To explicitly quantify the uncertainty associated with the phase prediction, we use a mixture density network (MDN) model. The architecture of our MDN model is shown in Figure 2. In this architecture, the hyperbolic tangent (tanh) activation function is used for all layers. It is important to note that the MDN model outputs a probability density function (PDF), parameterized by the mixture coefficients $(\pi_1, \ldots, \pi_m)$, means $(\mu_1, \ldots, \mu_m)$, and standard deviations $(\sigma_1, \ldots, \sigma_m)$, rather than a single deterministic value. This formulation enables the model to capture predictive uncertainty and to provide explicit confidence intervals for phase-fraction estimates.

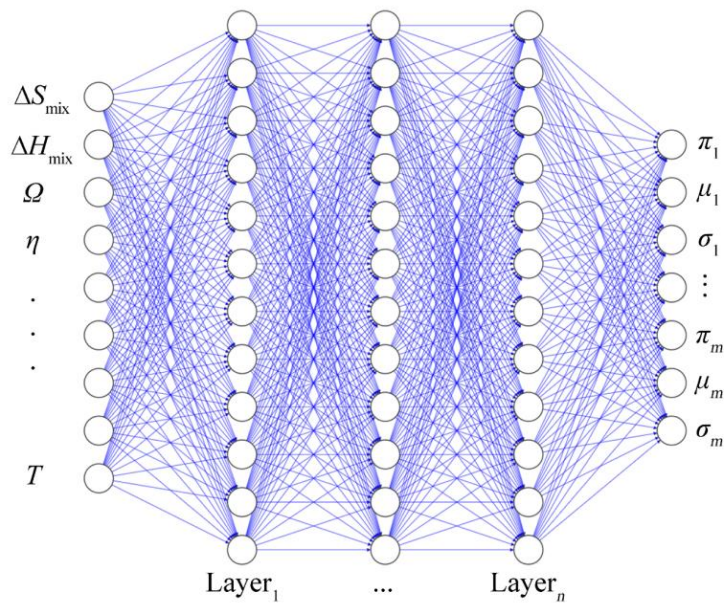

**Figure 2:** The architecture of the MDN network in this study. Note that *n* and *m* represent the number of hidden layers and the number of Gaussians, respectively.

Several parameters of our DL model, including the number of hidden layers, the number of neurons in each layer, batch size, the inclusion of batch normalization after the first layer, the inclusion of a dropout layer after the final layer, learning rate, and the number of Gaussians, are fine-tuned using Bayesian optimization [49,50]. More details on the Bayesian optimization procedure used for hyperparameter tuning are available in Supplementary Section II, while the corresponding results for the BCC phase are reported in Supplementary Table II. A summary of the optimized architecture is provided in Table . Importantly, the combined BCC/B2 phase is denoted simply as 'BCC' throughout the remainder of this paper.

**Table II:** The fine-tuned architecture for the different MDN networks trained in this study. Note that for Dropout (DO) and Batch Normalization (BN), a value of 0 indicates the layer is absent and a value of 1 indicates it is present, after layer$_1$ for BN and after layer$_n$ for DO.

| Network | Predicted phase | Batch size | BN | Drop-out | Layer number $n$ | Neurons | Learning rate | Gaussian number $m$ |
|---|---|---|---|---|---|---|---|---|
| $MDN_{FCC}$ | FCC | 699 | 0 | 0 | 7 | 55 | 0.00246 | 5 |
| $MDN_{BCC/B2}$ | BCC/B2 | 136 | 0 | 0 | 7 | 83 | 0.00108 | 4 |
| $MDN_{Laves}$ | Laves | 686 | 0 | 0 | 8 | 70 | 0.00275 | 3 |
| $MDN_{Sigma}$ | Sigma | 901 | 1 | 0 | 6 | 92 | 0.00069 | 5 |
| $MDN_{Heusler}$ | Heusler | 488 | 0 | 0 | 7 | 72 | 0.00173 | 5 |
| $MDN_{Liquid}$ | Liquid | 488 | 0 | 0 | 7 | 72 | 0.00173 | 5 |

We train the MDN by minimizing the negative log-likelihood (NLL) of the target values under the predicted Gaussian mixture. For each training sample, the MDN outputs $\pi_k$, $\mu_k$, and $\sigma_k$ for $k = 1, \ldots, K$ mixture components. The likelihood of the target y is computed as follows [51]:

$$p(y|\{\pi_k, \mu_k, \sigma_k\}) = \sum_{k=1}^{K} \pi_k N(y|\mu_k, \sigma_k) \quad (1)$$

Where $N(y|\mu_k, \sigma_k)$ is the Gaussian probability density function and is calculated as follows:

$$N(y|\mu, \sigma) = \frac{1}{\sigma\sqrt{2\pi}} \exp\left(-\frac{(y-\mu)^2}{2\sigma^2}\right) \quad (2)$$

The loss function is then calculated as follows:

$$L = -\frac{1}{N}\sum_{i=1}^{N} \log\left(\sum_{k=1}^{K} \pi_{ik} N(y_i|\mu_{ik}, \sigma_{ik})\right) \quad (3)$$

Where $N$ is the batch size. The parameters of the mixture, including mean $\mu$, standard deviations $\sigma$, and mixing coefficient of the Gaussian $\pi$, are predicted by the network and updated via backpropagation using the Adam optimizer, which adapts the learning rate during training for more efficient convergence [52].

## III. Results and Discussions

## A. Performance evaluation of the MDN model

To assess the predictive accuracy of the MDN model in phase prediction of RMPEAs, the prepared dataset is used to separately train six MDN models, one for each constituent phase. For training, 10% of the data (before oversampling) is randomly selected as the test set, with the remaining 90% used for training and validation. Within this subset, 10% is further set aside for validation, and the remaining 90% is oversampled following the procedure described in Section II A for the training process. Training stops at the epoch when the validation loss reaches a minimum. To evaluate the performance of the six MDN models after training, parity plots comparing the weighted predicted and true phase fractions for all six phases in the testing dataset are shown in Figure 3. For each data point, the predicted value corresponds to the weighted mean, and the total standard deviation of the mixture distribution quantifies the associated aleatoric uncertainty. These are calculated as:

$$\mu_{\text{weighted}} = \sum_{i=1}^{N} \pi_i \mu_i \tag{4}$$

$$\sigma_{\text{weighted}} = \sqrt{\sum_{i=1}^{N} \pi_i (\sigma_i^2 + (\mu_i - \mu_{\text{weighted}})^2)} \tag{5}$$

Where $N$ is the number of Gaussians. As is seen from Figure 3, the weighted mean values for most data points lie close to the 45° diagonal line, indicating high predictive accuracy across all phases. We further quantify on each figure the model's performance by reporting the coefficient of determination ($R^2$), computed using the weighted mean predictions for each phase. The consistently high $R^2$ values across all phases reinforce the model's accuracy and reliability in capturing phase distribution trends.

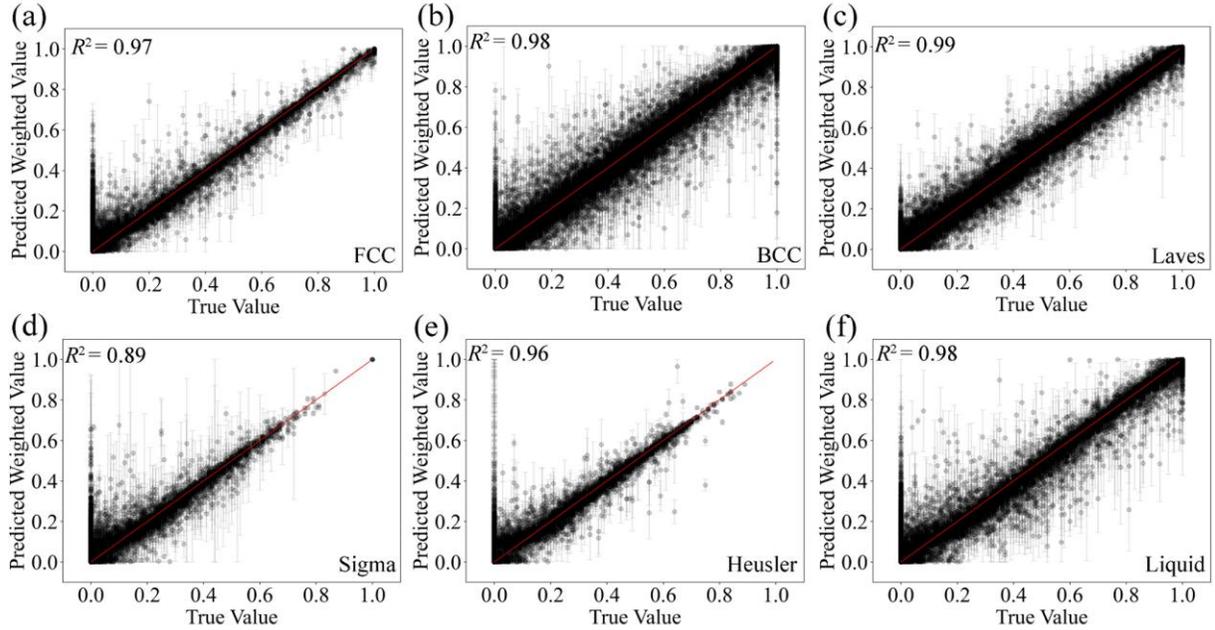

**Figure 3:** Parity plots of the predicted and true phase fraction of (a) FCC, (b) BCC, (c) Laves, (d) Sigma, (e) Heusler, and (f) Liquid phases for the testing dataset. Note that the mean and standard deviation values for each datapoint are computed using a weighted Gaussian mixture approach.

While the parity plots in Figure 3 quantify the overall model performance, it is difficult to infer from these plots a deeper understanding of the accuracy of the predicted fraction distributions for each phase. To address this, Figure 4 shows a 2D density mesh plot of true phase fractions versus prediction error, defined as the difference between the true and predicted weighted mean values for each phase in the testing dataset. The color intensity of each cell represents the number of RMPEAs associated with the corresponding predicted error versus the true values, thereby identifying regions of high and low error-prediction density and offering a more detailed view of the model's accuracy across the phase-fraction spectrum. Notably, despite employing a strict color scale limited to the range 0–5 for the density of points, most of the activated cells are concentrated along the horizontal line defined by prediction error = 0. This indicates that the model performs consistently well across the entire phase fraction range for all six phases.

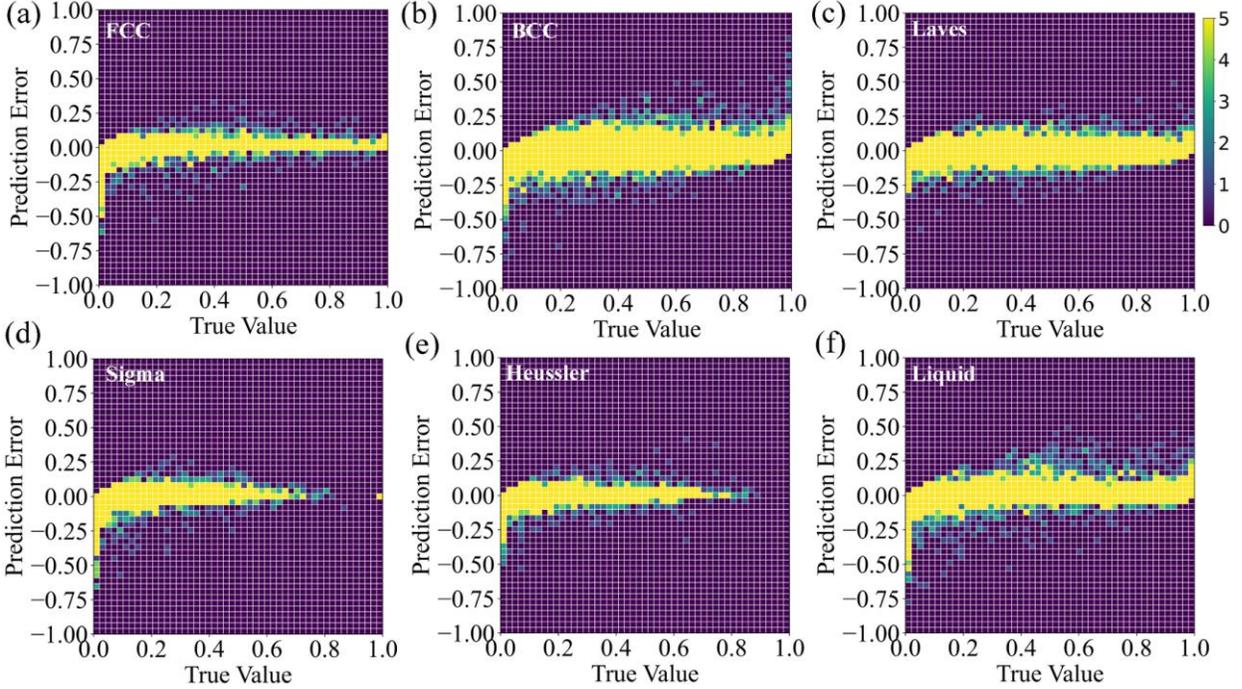

**Figure 4:** 2D histogram of the true-predicted and true phase fraction of (a) FCC, (b) BCC, (c) Laves, (d) Sigma, (e) Heusler, and (f) Liquid phases for the testing dataset, visualized as a colored mesh plot. The color intensity represents the density of the data points, with the color scale ranging from 0 to 5. Cells near True = 1 are not activated for the Heusler phase because our dataset does not include any single-phase Heusler samples.

As a representative case, Figure 5(a) shows the predicted probability density from our MDN for the BCC phase fraction of $Ti_{0.32}Al_{0.26}V_{0.19}Nb_{0.11}Co_{0.12}$ at 974 K, where the true label is 0.91. The MDN outputs a Gaussian mixture representing the distribution of possible phase fractions, from which we compute the mixture mean and variance. For this composition, the mean predicted value is 0.95, and the corresponding aleatoric uncertainty, which is quantified as the standard deviation of the mixture, is 0.05. This uncertainty reflects the intrinsic variability in the model's prediction arising from ambiguity in mapping the input features to the output phase fraction. As a result, although the true value is a single deterministic number, the MDN predicts a probability distribution that encodes the range of plausible outcomes due to aleatoric effects.

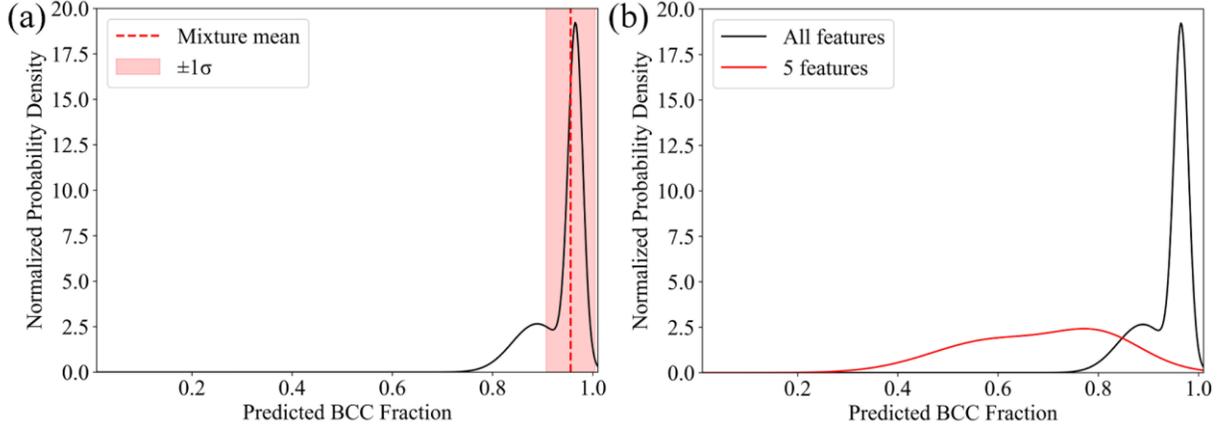

**Figure 5:** (a) MDN-predicted probability density of the BCC phase fraction for $Ti_{0.32}Al_{0.26}V_{0.19}Nb_{0.11}Co_{0.12}$ with the true label of 0.91 at T = 947 K, wherein shading represents aleatoric uncertainty. (b) MDN-predicted probability density for the same composition using a subset of the feature set, highlighting epistemic uncertainty that can arise from a reduction in feature inputs.

## B. Feature importance and epistemic uncertainty

As discussed in the earlier section, we have successfully trained six MDN models capable of predicting the phase fractions of RMPEAs, while quantifying the corresponding aleatoric uncertainty. A critical next step is to evaluate epistemic uncertainty arising from incomplete knowledge of the optimal set of input features and determine the minimum number of features required for accurate phase prediction among the 41 features. To this end, we perform a perturbation-based feature importance analysis [53,54] on the testing dataset using the trained models to assess the contribution of each input feature to the model's predictive performance. We focus this analysis on the BCC phase due to its central role in enhancing the strength of RMPEAs [55].

The feature importance ranking obtained from the perturbation analysis is presented in Figure 6(a). The top five features identified by the MDN model during training are the Covalent radius, Crystal radius, MB electronegativity, $\Omega$, and atomic size mismatch $\delta$. To evaluate the predictive capacity of a reduced feature set and the corresponding epistemic uncertainty, we retrain the MDN model using only these five features, keeping the network architecture unchanged. However, when evaluated on the testing dataset, the model exhibits significantly degraded performance, with $R^2$ of only 0.63, and a parity plot indicating notable deviation from the $45^o$ line shown in Figure 6(b). This degradation reflects heightened epistemic uncertainty introduced by insufficient input knowledge, underscoring that the five top-ranked features alone do not capture the full complexity of BCC phase stability in RMPEAs. As a representative case, Figure 5(b) shows the predicted probability density of the BCC phase fraction for $Ti_{0.32}Al_{0.26}V_{0.19}Nb_{0.11}Co_{0.12}$ at 974 K, where the true label is 0.91, for two MDN models: one trained using all 41 input features and one trained using only the top 5 features. Interestingly, while the mean prediction decreases moderately from 0.95 to 0.71, the predicted uncertainty increases substantially from 0.05 in the full-feature model to 0.18 in the reduced-feature model, illustrating the pronounced impact of epistemic uncertainty on the predicted distribution. This increase in uncertainty arises because the reduced-feature model lacks some features necessary to successfully predict the RMPEA phases, limiting the model's knowledge and confidence in its predictions. Unlike aleatoric uncertainty, which reflects the model's ambiguity in mapping input features to the output phase fraction, epistemic uncertainty can be reduced by incorporating additional relevant features.

To determine the minimum number of features required for reliable phase prediction, we incrementally expand the input feature set based on the importance ranking and retrain the model at each step. Notably, when the model is trained on the top 12 features, performance improves markedly, achieving an $R^2$ of 0.98 on the test set. The parity plot shown in Figure 6(c) confirms excellent agreement between the weighted mean predicted and true values, indicating that these 12 features capture the essential information governing phase formation in RMPEAs. This reduction in error, along with the improvement in $R^2$, indicates a substantial decrease in epistemic uncertainty, suggesting that the inclusion of additional informative features compensates for earlier deficiencies in feature representation.

Beyond 12 features, we observe no significant gain in predictive performance with the $R^2$ values plateauing, as illustrated in Figure 6(d). This saturation suggests that the remaining features contribute little additional value and may introduce noise or redundancy. Therefore, we conclude that a subset of 12 carefully selected features is sufficient for accurate phase prediction, achieving performance comparable to models trained on all 41 features. This optimal subset not only minimizes epistemic uncertainty but also enhances model interpretability and robustness by eliminating irrelevant or redundant inputs that may obscure the underlying physics. Our findings highlight the importance of systematically addressing epistemic uncertainty when developing predictive models for complex materials systems.

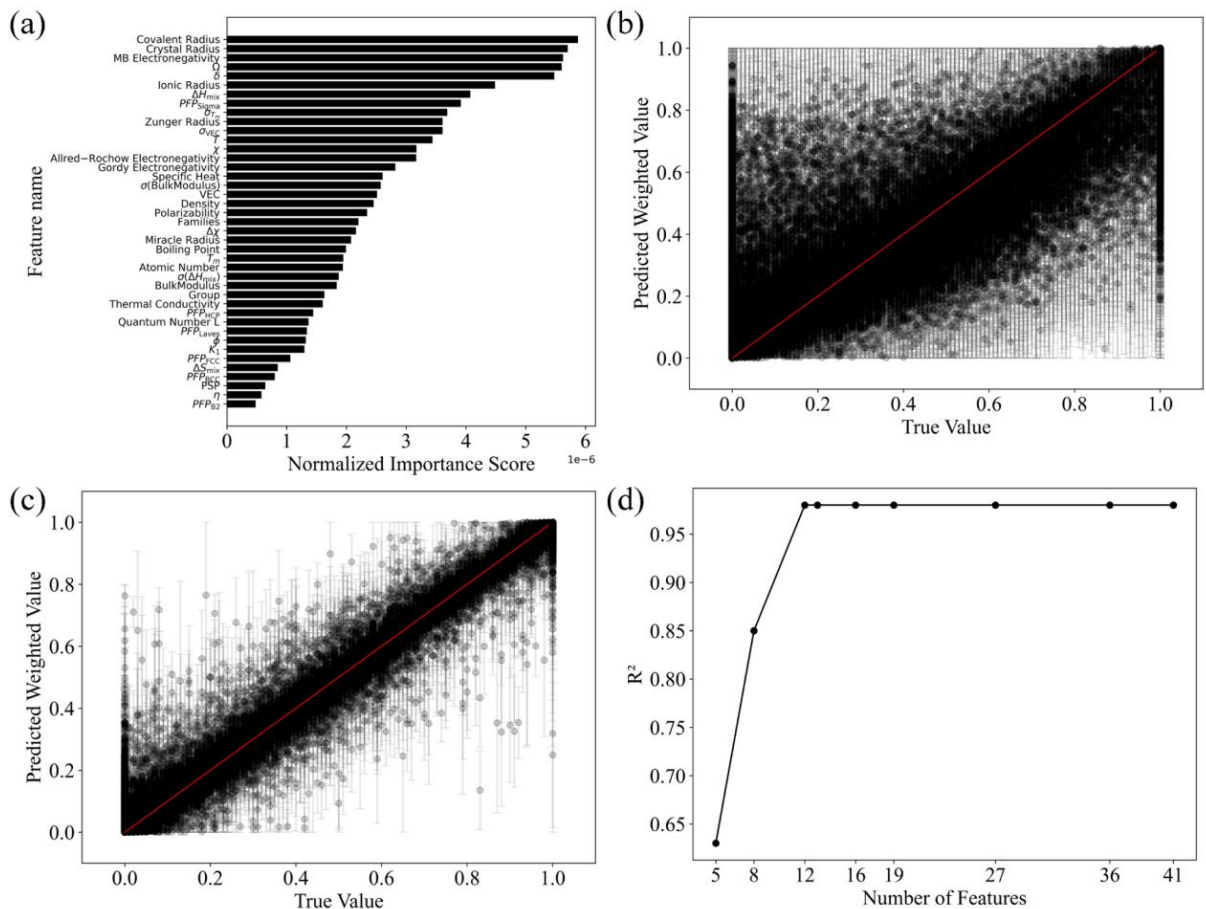

**Figure 6:** (a) Ranked importance of input features for BCC phase prediction. (b-c) Parity plots comparing the predicted and true BCC phase fractions using five and twelve input features, respectively. (d) Variation of the $R^2$ with the number of input features used for the BCC phase prediction.

## C. Out-of-distribution (OOD) Discovery of RAMPEs

In this section, we propose an uncertainty-aware active learning approach to discover novel RMPEAs with the target phase that incorporates previously unseen elements. Conventional deep learning models such as MLPs often produce overconfident predictions when extrapolating to unfamiliar regions of the design space. In contrast, MDNs provide explicit uncertainty quantification, enabling the model to distinguish between high-confidence predictions and regions where reliability is low. As a result, while a conventional MLP trained on known alloys might suggest candidates in a new elemental space, it cannot distinguish between meaningful extrapolations and unreliable guesses. This distinction is particularly important in out-of-distribution (OOD) scenarios, where the design space differs from the training domain, and relying solely on point predictions without uncertainty can lead to wasted experimental effort on compositions unlikely to yield the target phase. Our uncertainty-guided strategy thus ensures that extrapolation of unexplored design space is both efficient and informed, addressing the limitations of standard models that do not account for uncertainty.

Our setup reflects a practical discovery scenario in ML-assisted high-throughput experimentation, where researchers aim to identify materials with desired properties in composition spaces that differ significantly from the training domain. For instance, a model may be tasked with proposing 100 candidate alloys predicted to form the target phase. If 80 of these are experimentally confirmed, the resulting data can be leveraged to fine-tune the model and enhance its extrapolation capability. Iterative feedback is essential when exploring previously uncharted chemical spaces, as it guides discovery and improves model robustness.

As in the earlier section, we maintain our focus on the BCC phase due to its critical role in strength. To start with, we reorganize our dataset so that all Ti-free alloys are included in the training set, while all Ti alloys form the testing dataset. We then train an MDN model to predict the BCC phase fraction and its associated aleatoric uncertainty using the updated dataset. The model exhibits poor performance on the testing dataset, as illustrated in Supplementary Figure S2. This outcome is expected, given that ML models typically struggle to generalize to out-of-distribution data. To enable the discovery of BCC Ti-alloys, we implement a two-stage uncertainty-based active learning approach as follows: 1) We begin by evaluating all compositions in the testing dataset using the MDN model trained on Ti-free alloys. From this set, we identify all alloys predicted to exhibit a BCC phase fraction of at least 90%. 2) The selected candidates are clustered into a predefined number of groups (initially 500) using k-means clustering. 3) From each cluster, the sample closest to the cluster centroid is chosen, resulting in 500 samples referred to as the "selected Ti-alloy candidates." Of these, 90% are randomly added to the training dataset, while the remaining 10% serve as a validation dataset. 4) The training process is continued, and it is stopped at the epoch where the validation loss reaches its minimum. 5) Steps 1 through 4 are repeated for a predefined number of active-learning cycles (three cycles in this study). After these cycles, step 1 is modified as follows: instead of selecting all alloys predicted to exceed the 90% BCC threshold, we first sort the alloys by model uncertainty and select a predefined number (three times the desired number of clusters). 6) We then proceed with steps 2 through 4 using this updated selection strategy, and we repeat the active-learning cycle accordingly. The reason for not focusing on the uncertainty in the initial stage is that the model has not yet been exposed to any Ti-alloys, making its uncertainty estimates unreliable for guiding sample selection. Importantly, our active learning strategy is inspired by the framework proposed by Doucet et al [56].

Figure 7(a) shows the BCC fraction of the 500 selected Ti-alloy candidates obtained in each selection round, which are then added to the training dataset for the subsequent active learning cycle. As a result, the first point represents the BCC fraction of the 500 Ti-alloy candidates selected by the original MDN model and added to the training dataset to initiate the first active learning cycle. For comparison, we report the performance of an alternative strategy in which alloys with the highest predicted uncertainty are chosen instead of those with the lowest in step 5 of the active learning. It is important to note that each cycle suggests 500 new Ti-alloy candidates that differ from those in earlier cycles. This is because the 500 selected Ti-alloy candidates are removed from the testing dataset at each cycle, and subsequent selections are made from the remaining samples. As observed from Figure 7(a), the initial BCC fraction is relatively low. Interestingly, the BCC fraction rises during the first three cycle selections (i.e., first stage of active learning). During the second stage of active learning, the BCC fraction increases rapidly, then plateaus at approximately 0.9 for the low-uncertainty route. In contrast, it decreases and plateaus at approximately 0.6 for the high-uncertainty route. These results demonstrate that using the low-uncertainty route, our active learning strategy enables the MDN model to consistently identify 500 novel BCC Ti-alloys with high accuracy in each cycle.

To further evaluate the performance of the MDN model throughout the active learning process, the Precision, Recall, and F1 score on the remaining testing dataset at different cycles are reported in Figure 7(b–d) for both active learning routes. Notably, the high-uncertainty route consistently achieves the highest precision, exceeding 0.75 after two cycles, as shown in Figure 7(b). In contrast, the recall of the low-uncertainty route fluctuates around 0.3, whereas the recall of the high-uncertainty route steadily increases over the cycles, as seen in Figure 7(c). As a result, the high-uncertainty route exhibits continuous improvement in the F1 score shown in Figure 7(d) while the F1 score for the low-uncertainty route shows only minimal change.

These trends highlight a clear exploitation–exploration trade-off between the two active learning routes. The low-uncertainty route primarily exploits regions of the design space where the MDN model is already confident. As a result, the selected candidates are enriched in true BCC alloys in every cycle, leading to the rapid increase in BCC fraction observed in Figure 7(a). This exploitation-driven approach yields minimal improvements in F1 score. In contrast, the high-uncertainty route emphasizes exploration by selecting alloys from regions where the model is highly uncertain. Because these regions contain a broader mix of phases, the BCC fraction of the selected samples is lower and plateaus around 0.6. However, this exploratory behavior expands the model's coverage of the design space of Ti-alloys, enabling it to discover more true BCC alloys over time. This effect is reflected in the steadily increasing recall and, thereby, the F1 score. Thus, while exploitation provides high-accuracy candidates in each cycle through a low-uncertainty route, exploration enhances the model's long-term ability to discover new BCC regions through a high-uncertainty route.

Together, these findings confirm that the low-uncertainty route is particularly well-suited for exploitation-driven applications, such as the targeted discovery scenario introduced at the beginning of this section. This scenario focuses on the autonomous identification of BCC RMPEAs to guide experimentalists. In contrast, a low-uncertainty route may be more effective when the goal is broader exploration across the design space.

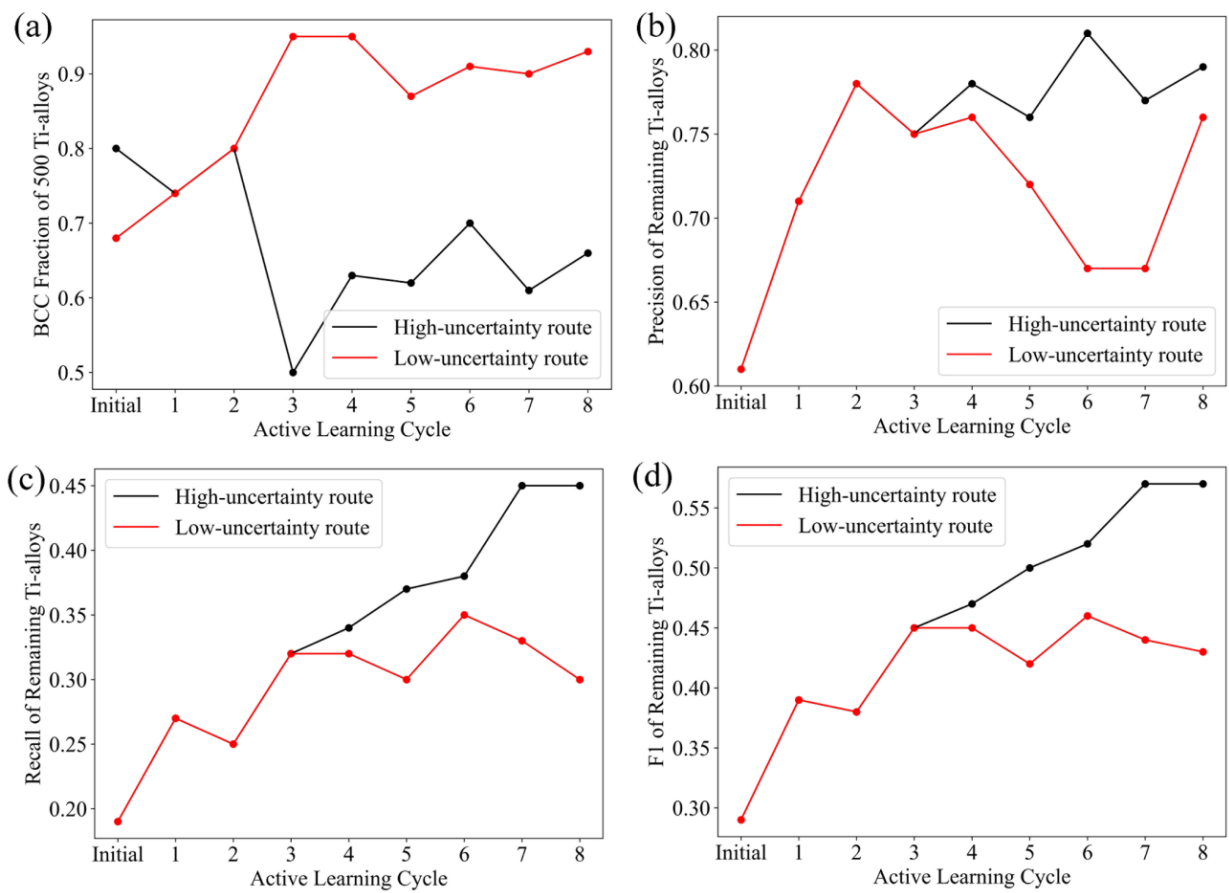

**Figure 7:** (a) BCC fraction of the 500 selected Ti-alloy candidates in each active learning cycle. (b) Precision of the remaining Ti-alloys from the testing dataset in each active learning cycle. (c) Recall of the remaining Ti-alloys from the testing dataset in each active learning cycle. (d) F1 of the remaining Ti-alloys from the testing dataset in each active learning cycle. These opposing trends underscore the fundamental trade-off between exploitation and exploration.

Next, we examine how the acquisition size of the selected Ti-alloy candidates influences the discovery of BCC Ti-alloys through the low-uncertainty route, as shown in Figure 8. It is seen from Figure 8(a) that although the initial BCC fraction of the selected candidates is low for all acquisition sizes, the BCC fraction increases and stabilizes above 0.8 after three active learning cycles for sample sizes of 100 or larger. In contrast, the trends for acquisition sizes of 30 and 50 fluctuate and do not exhibit consistent improvement possibly due to limited coverage of the compositional space in each cycle, which reduces the statistical reliability of the selected candidates.

Notably, Figure 8(b) reveals that while the F1 score is generally considered low across all sample sizes, it exhibits a relatively larger improvement over successive active learning cycles for the largest acquisition size than for the smaller sizes. In contrast, smaller acquisition sizes result in F1 values that fluctuate around 0.33. This observation implies that larger acquisition sizes enable the model to capture more informative and diverse samples in each cycle, providing a more comprehensive view of the feature distribution patterns for BCC samples within the Ti-alloy design space.

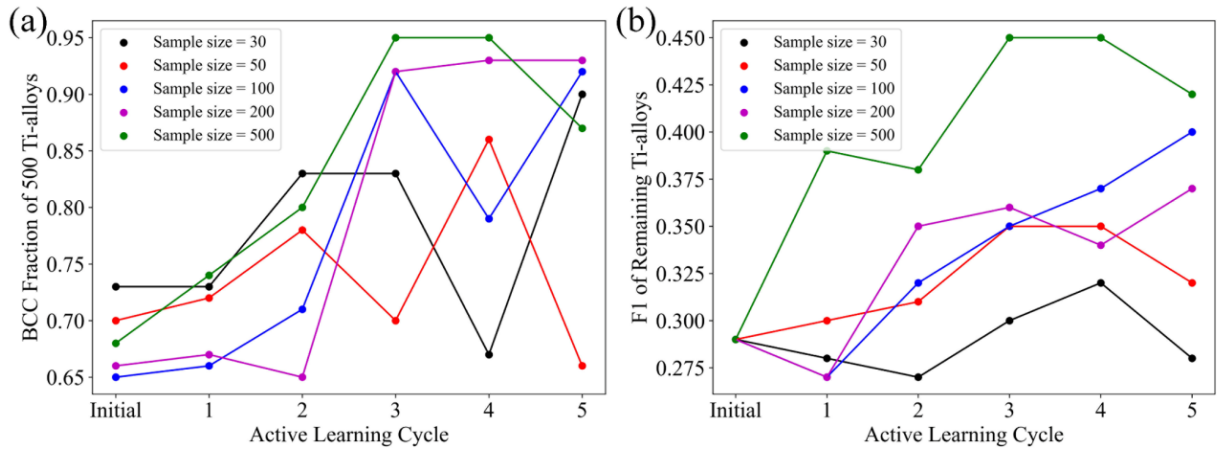

**Figure 8:** (a) BCC Fraction of the selected Ti-alloy candidates in each active learning cycle for varying acquisition sizes. (b) F1 of the remaining Ti-alloys from the testing dataset in each cycle as a function of acquisition size.

A final concern is whether the selected Ti-alloy candidates are concentrated in a narrow region of the design space. To investigate this, Figure 9 shows violin plots representing the distribution of elemental fractions for BCC compositions in the original Ti-alloy dataset and those discovered through the active learning cycles from low-uncertainty route. It is seen from the figure that the range of elemental compositions for the selected Ti-alloy candidates across all active learning cycles closely resembles that of the original BCC dataset. This indicates that the uncertainty-based active learning strategy does not confine discovery to a narrow region but instead identifies compositions broadly distributed across the design space.

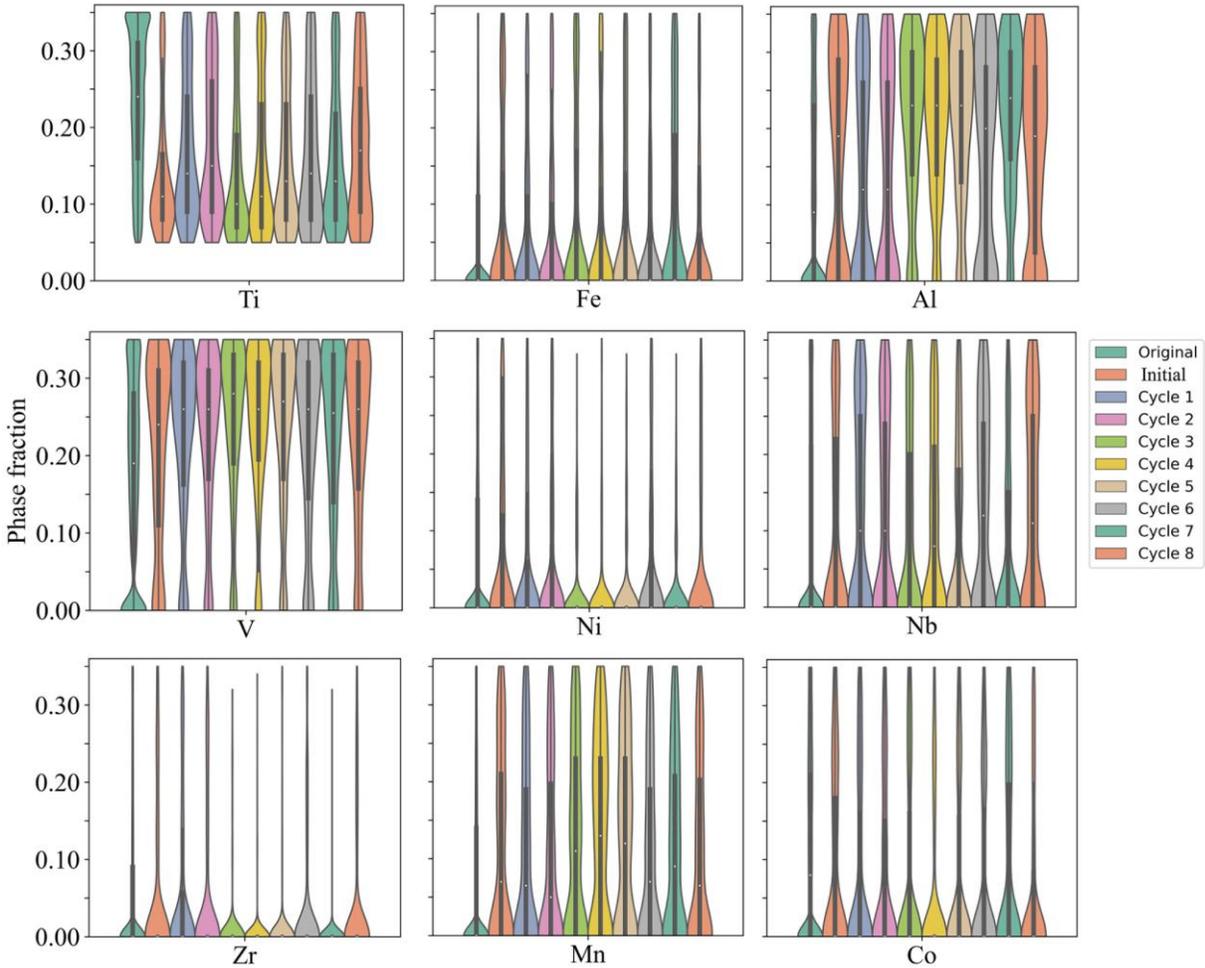

**Figure 9**: Distribution of elemental fractions for compositions in the original Ti-alloy dataset and selected Ti-alloy candidates during eight active learning cycles (Cycle 1 to Cycle 8). The elements are grouped for clarity.

## IV. Conclusion

Refractory multi-principal element alloys (RMPEAs) represent a distinct class of metallic alloys with an expansive compositional design space and exceptional mechanical performance under extreme conditions. Phase stability has been shown to play a critical role in governing the mechanical properties of RMPEAs, motivating the development of data-driven surrogate models to predict their phase formation. However, most existing machine learning models have focused on learning a deterministic mapping from composition-derived features to expected phase labels, without accounting for the uncertainty associated with their predictions. This limitation can result in excessively simplistic models that are overly confident and tend to generalize poorly when applied to compositions outside the training domain.

In this study, we developed a deep learning framework based on a Mixture Density Network (MDN) to accurately predict the phase fractions of RMPEAs and quantify the associated uncertainty across a range of temperatures. The key conclusions from our study are:

- Our training dataset was constructed by randomly sampling 70,000 compositions from the design space spanned by Ti, Fe, Al, V, Ni, Nb, Zr, Mn, and Co. Each composition was then labeled with the expected phase fractions at seven different temperatures using CALPHAD. Due to a significant class imbalance across the phases, we applied an oversampling strategy to balance the training data prior to training. Using the oversampled dataset, we successfully trained six separate MDN models, each corresponding to one target phase: FCC, BCC, Laves, Sigma, Heusler, and Liquid. These models predicted phase fractions and their associated aleatoric uncertainty with high accuracy in a 41-dimensional feature space. The $R^2$ values (0.97 for FCC, 0.98 for BCC, 0.99 for Laves, 0.89 for Sigma, 0.96 for Heusler, and 0.98 for Liquid) confirmed the high accuracy and robustness of our framework in capturing phase distribution trends in RMPEAs.

- A common limitation in previous studies was the inconsistent selection of input features, reflecting incomplete knowledge about the most informative descriptors for phase prediction. This variability highlighted the need to quantify epistemic uncertainty arising from such incomplete feature knowledge. To address this, we performed a perturbation-based feature importance analysis focused on the BCC phase. Our analysis showed that reducing the input feature set from 41 to the 12 most important features preserved both predictive accuracy and uncertainty levels. However, further reduction beyond this point led to noticeable drops in accuracy and increased epistemic uncertainty, emphasizing the importance of selecting a minimally sufficient yet informative feature subset.
- We further proposed an active learning framework to discover novel RMPEAs with the BCC phase, incorporating previously unseen elements. Using Ti-free alloys for training and Ti-alloys as prediction candidates, we compared two acquisition strategies: low-uncertainty and high-uncertainty routes. Our results showed that low-uncertainty sampling rapidly improved the BCC fraction of model-suggested Ti-alloys from 0.68 to nearly 0.9, while high-uncertainty sampling showed minimal improvement.
- Our analysis highlighted a clear exploitation–exploration trade-off between the two active learning routes. The low-uncertainty route primarily exploited regions of the design space where the MDN model was already confident. As a result, while the selected candidates were enriched in true BCC alloys in every cycle, it yielded minimal improvements in the F1 score. In contrast, the high-uncertainty route emphasized exploration by selecting alloys from regions where the model was highly uncertain. Because these regions contained a broader mix of phases, the BCC fraction of the selected samples was lower. However, this exploratory behavior expanded the model's coverage of the design space of Ti-alloys, steadily increasing in recall and thereby F1 score. Our calculations further revealed that the model-suggested Ti-alloys span a broad region of the design space, rather than being confined to a narrow subset.

- The data-driven framework developed in this study is applicable across a wide range of technologically relevant temperatures and adaptable to various alloy systems. By explicitly quantifying both aleatoric and epistemic uncertainties, it serves as a robust and broad transferable tool for accurate phase prediction and the rapid discovery of novel materials with reliable confidence estimates.

# Acknowledgment

CAIMEE research was sponsored by the Army Research Laboratory and was accomplished under Cooperative Agreement Number W911NF-22-2-0014. The authors gratefully acknowledge internal financial support from the Johns Hopkins Applied Physics Laboratory's Independent Research & Development (IR&D) Program. Computational resources were provided by the Advanced Research Computing at Hopkins (ARCH).

# Declaration of Competing Interest

The authors declare that they have no known competing financial interests or personal relationships that could have appeared to influence the study reported in this paper.

# Author contributions

**Ali K. Shargh**: Writing – original draft, Writing – review and editing, Visualization, Validation, Software, Methodology, Investigation, Formal analysis, Data curation, Conceptualization.

**Christopher D. Stiles**: Writing – review and editing, Methodology, Funding acquisition, Formal analysis, Conceptualization.

**Jaafar A. El-Awady**: Writing – review and editing, Supervision, Resources, Project administration, Methodology, Investigation, Funding acquisition, Formal analysis, Conceptualization.

# Supplementary Material for

# Uncertainty-aware phase fraction prediction and active-learning-guided out-of-domain discovery of refractory multi-principal element alloys

Ali K. Shargh[1*], Christopher D. Stiles[1,2], Jaafar A. El-Awady[1†]

[1] Department of Mechanical Engineering, Johns Hopkins University, Baltimore, Maryland 21218, United States

[2] Research and Exploratory Development Department, Johns Hopkins Applied Physics Laboratory, Laurel, Maryland 20723, United States

## I. Evaluation of sampling strategies and phase fraction predictions

We provide the distributions of RMPEA phase fractions of the original dataset, excluding the testing dataset, after oversampling in Supplementary Figure S1. Comparison of MDN model performance for the FCC phase using different sampling strategies on the testing set is reported in Table S1. In addition, the Parity plots of the predicted and true BCC phase fraction for the Ti-alloys using the network trained on Ti-free alloys and a 2D histogram of the predicted and true BCC fraction for the Ti-alloys using the network trained on Ti-free alloys, visualized as a colored mesh plot, are provided in Figure S2.

---

[*] Contact author: ashargh1@jhu.edu (A. K. Shargh)
[†] Contact author: jelawady@jhu.edu (J. A. El-Awady)

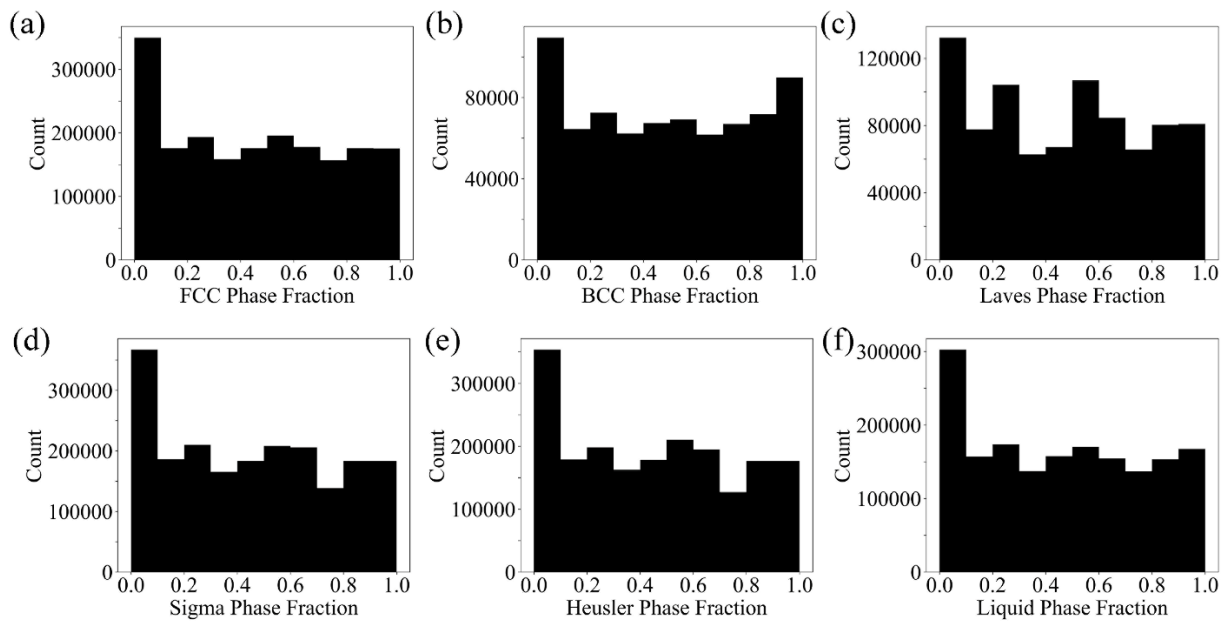

**Figure S1**: Distributions of RMPEA phase fractions of the original dataset training dataset after oversampling.

**Table I: Comparison of MDN model performance for the FCC phase using different sampling strategies on the testing set.**

| ID | Approach | Loss | $R^2$ |
|---|---|---|---|
| 1 | Balanced Random Oversampling | -7.54 | 0.98 |
| 2 | Extreme Random Oversampling | -7.43 | 0.95 |
| 3 | SMOTER | -7.54 | 0.95 |
| 4 | Gaussian Noise | -7.52 | 0.98 |
| 5 | WERCS | -4.91 | 0.95 |

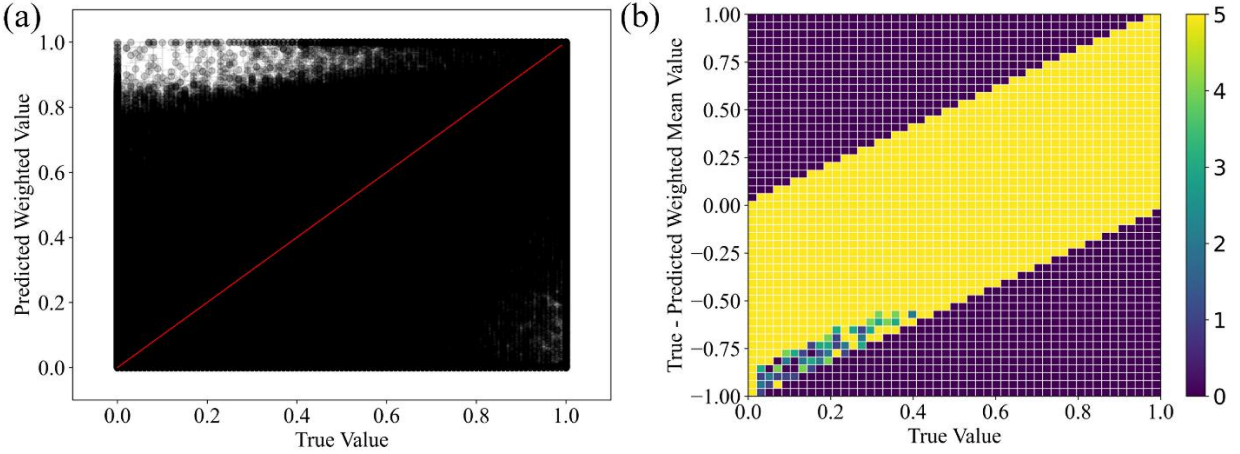

**Figure S2:** (a) Parity plots of the predicted and true BCC phase fraction for the Ti-alloys using the network trained on Ti-free alloys. (b) 2D histogram of the predicted and true BCC fraction for the Ti-alloys using the network trained on Ti-free alloys, visualized as a colored mesh plot.

## II. Bayesian optimization for neural network tuning

To optimize the neural network architecture for predicting mechanical properties of RMPEAs, we employ Bayesian Optimization. The search space includes batch size (32–1024), number of neurons per layer (20–100), number of layers before the dropout layer (5–10), number of Gaussians (2-6), and learning rate (0.0001-0.01). The objective function is to minimize the mean square error (MSE), evaluated on a held-out test set. We run 40 random initial evaluations followed by 80 iterations of optimization. This process improves the model's performance. As a representative case, we report the results of the Bayesian optimization for the BCC phase in Table S2.

**Table II:** The results of our Bayesian optimization for the BCC phase. Note that a value of 0 or 1 for Dropout (DO) and batch normalization (BN) layers indicates their absence or presence, respectively.

| DO | DO rate | BN | Batch size | L1 | Neurons | Learning rate | Gaussians | Loss |
|---|---|---|---|---|---|---|---|---|
| 0 | 0.31803 | 0 | 639 | 9 | 32 | 0.00302 | 4 | -3.34338 |
| 0 | 0.59536 | 1 | 269 | 6 | 74 | 0.0009 | 3 | -3.4517 |
| 0 | 0.13698 | 1 | 494 | 10 | 87 | 0.00796 | 6 | -2.27859 |
| 1 | 0.31066 | 1 | 605 | 5 | 28 | 0.0046 | 5 | -3.05637 |
| 0 | 0.59924 | 0 | 593 | 6 | 59 | 0.00619 | 5 | -3.25777 |
| 0 | 0.25948 | 0 | 352 | 10 | 85 | 0.0092 | 6 | -1.60453 |
| 1 | 0.34064 | 0 | 975 | 10 | 46 | 0.00423 | 2 | -2.42346 |
| 0 | 0.11669 | 0 | 85 | 6 | 34 | 0.00548 | 3 | -2.89972 |
| 0 | 0.48412 | 0 | 710 | 8 | 68 | 0.00447 | 6 | -3.34087 |
| 1 | 0.34896 | 1 | 179 | 8 | 34 | 0.00688 | 4 | -2.7607 |
| 0 | 0.28075 | 1 | 101 | 7 | 34 | 0.00188 | 5 | -3.24808 |
| 1 | 0.5209 | 1 | 166 | 5 | 90 | 0.00212 | 3 | -3.27363 |
| 1 | 0.20525 | 0 | 761 | 9 | 45 | 0.00069 | 6 | -3.18776 |
| 1 | 0.15759 | 1 | 934 | 10 | 70 | 0.00101 | 3 | -3.41078 |
| 1 | 0.28699 | 0 | 680 | 9 | 70 | 0.00897 | 3 | -1.58884 |
| 0 | 0.36612 | 0 | 663 | 7 | 58 | 0.00663 | 2 | -2.27789 |
| 0 | 0.27337 | 0 | 441 | 10 | 75 | 0.00223 | 5 | -3.53952 |
| 1 | 0.39591 | 0 | 728 | 6 | 22 | 0.00722 | 3 | -2.94947 |
| 0 | 0.52029 | 1 | 137 | 6 | 39 | 0.00987 | 5 | -2.56357 |
| 1 | 0.23265 | 1 | 109 | 10 | 52 | 0.0082 | 3 | -0.92602 |
| 0 | 0.56292 | 0 | 550 | 9 | 29 | 0.00506 | 6 | -3.28422 |
| 1 | 0.14742 | 1 | 540 | 9 | 71 | 0.00072 | 4 | -3.40899 |
| 0 | 0.54493 | 0 | 686 | 10 | 71 | 0.00741 | 4 | -1.59131 |
| 1 | 0.23344 | 1 | 298 | 10 | 44 | 0.00221 | 2 | -3.05404 |
| 1 | 0.34913 | 1 | 62 | 6 | 88 | 0.00019 | 5 | -3.36701 |
| 1 | 0.2941 | 1 | 688 | 7 | 58 | 0.0083 | 2 | -2.66766 |
| 0 | 0.15904 | 0 | 426 | 6 | 94 | 0.0075 | 4 | -2.85139 |
| 1 | 0.11249 | 1 | 898 | 9 | 61 | 0.00609 | 5 | -2.84011 |
| 0 | 0.22909 | 1 | 204 | 5 | 48 | 0.00557 | 4 | -3.14764 |
| 0 | 0.23676 | 0 | 136 | 9 | 84 | 0.00399 | 2 | -2.02762 |
| 0 | 0.43142 | 1 | 43 | 8 | 59 | 0.00584 | 5 | -2.03364 |
| 0 | 0.57007 | 1 | 807 | 9 | 95 | 0.00703 | 4 | -2.55847 |
| 0 | 0.38967 | 1 | 374 | 5 | 71 | 0.00557 | 3 | -3.23487 |
| 0 | 0.22574 | 1 | 604 | 10 | 30 | 0.00898 | 6 | -2.71118 |
| 0 | 0.38857 | 0 | 488 | 7 | 72 | 0.00173 | 5 | -3.6292 |
| 1 | 0.46465 | 1 | 925 | 6 | 40 | 0.00423 | 3 | -3.16297 |
| 1 | 0.50653 | 0 | 478 | 5 | 66 | 0.00264 | 5 | -3.25677 |
| 0 | 0.48221 | 1 | 557 | 6 | 20 | 0.00382 | 5 | -3.04733 |
| 1 | 0.25742 | 1 | 125 | 6 | 79 | 0.00272 | 2 | -3.02488 |
| 1 | 0.30266 | 0 | 44 | 9 | 87 | 0.00088 | 4 | -3.36903 |
| 1 | 0.22388 | 1 | 114 | 10 | 54 | 0.01 | 2 | 0.32607 |
| 1 | 0.24208 | 0 | 113 | 8 | 57 | 0.00603 | 2 | -2.03596 |
| 0 | 0.53729 | 1 | 116 | 8 | 52 | 0.0091 | 3 | -2.07229 |
| 0 | 0.56708 | 1 | 109 | 7 | 54 | 0.00068 | 2 | -3.31483 |
| 0 | 0.49106 | 0 | 109 | 8 | 49 | 0.00818 | 4 | -2.10185 |
| 1 | 0.1661 | 1 | 113 | 10 | 54 | 0.01 | 2 | -0.29912 |
| 1 | 0.57235 | 1 | 114 | 8 | 54 | 0.00803 | 5 | -1.99084 |
| 1 | 0.40562 | 0 | 106 | 10 | 52 | 0.00603 | 2 | -0.21927 |
| 1 | 0.1 | 1 | 114 | 10 | 54 | 0.01 | 2 | 0.28659 |
| 1 | 0.1 | 0 | 107 | 10 | 51 | 0.01 | 2 | 0.32538 |
| 1 | 0.38978 | 0 | 106 | 9 | 52 | 0.00981 | 3 | -1.60949 |
| 1 | 0.33331 | 0 | 107 | 9 | 51 | 0.0077 | 3 | -1.58852 |
| 1 | 0.28956 | 1 | 115 | 9 | 57 | 0.00902 | 2 | 0.03487 |
| 0 | 0.4645 | 0 | 116 | 9 | 56 | 0.00167 | 3 | -3.42544 |
| 1 | 0.13889 | 0 | 685 | 9 | 72 | 0.00125 | 4 | -3.53082 |
| 1 | 0.56823 | 1 | 113 | 8 | 55 | 0.00415 | 4 | -2.71837 |
| 1 | 0.52485 | 1 | 110 | 9 | 51 | 0.00292 | 4 | -2.81486 |
| 0 | 0.50091 | 0 | 105 | 10 | 53 | 0.00028 | 4 | -3.57873 |
| 1 | 0.58893 | 1 | 108 | 8 | 51 | 0.00426 | 3 | -2.67005 |
| 0 | 0.39109 | 1 | 108 | 9 | 49 | 0.00293 | 2 | -3.00035 |
| 1 | 0.1914 | 0 | 679 | 7 | 69 | 0.00503 | 2 | -2.1716 |
| 1 | 0.23827 | 0 | 109 | 9 | 53 | 0.00133 | 4 | -3.39017 |
| 1 | 0.35203 | 1 | 114 | 10 | 57 | 0.00313 | 4 | -2.85727 |
| 0 | 0.30822 | 1 | 115 | 8 | 56 | 0.0038 | 2 | -2.66945 |
| 1 | 0.17816 | 1 | 113 | 8 | 52 | 0.00456 | 3 | -2.78391 |
| 1 | 0.54626 | 0 | 679 | 7 | 71 | 0.00898 | 5 | -1.61531 |
| 1 | 0.4736 | 0 | 113 | 10 | 51 | 0.00405 | 3 | -2.5057 |
| 1 | 0.54154 | 0 | 115 | 10 | 58 | 0.00143 | 2 | -3.13294 |
| 0 | 0.30382 | 0 | 680 | 8 | 72 | 0.00332 | 4 | -3.52163 |
| 0 | 0.54904 | 0 | 105 | 9 | 50 | 0.00813 | 2 | 0.33264 |
| 0 | 0.47634 | 0 | 114 | 10 | 56 | 0.00104 | 2 | -3.2735 |
| 0 | 0.42301 | 0 | 113 | 10 | 53 | 0.00881 | 5 | -1.58872 |
| 1 | 0.53002 | 0 | 114 | 9 | 55 | 0.00235 | 3 | -2.97297 |
| 0 | 0.30546 | 1 | 106 | 10 | 50 | 0.00365 | 3 | -2.78179 |
| 0 | 0.29958 | 0 | 109 | 8 | 50 | 0.00216 | 5 | -3.35551 |
| 1 | 0.13122 | 1 | 114 | 10 | 54 | 0.01 | 2 | 0.33311 |
| 0 | 0.13346 | 1 | 110 | 10 | 51 | 0.00961 | 2 | 0.09862 |
| 1 | 0.34465 | 0 | 105 | 9 | 51 | 0.0037 | 3 | -2.55715 |
| 1 | 0.57326 | 0 | 679 | 10 | 69 | 0.00342 | 3 | -3.32725 |
| 0 | 0.55657 | 1 | 104 | 9 | 51 | 0.00995 | 3 | -0.73332 |
| 0 | 0.3837 | 1 | 113 | 9 | 55 | 0.00484 | 3 | -2.55533 |
| 1 | 0.28314 | 1 | 112 | 10 | 53 | 0.00657 | 6 | -1.30271 |
| 1 | 0.14221 | 1 | 111 | 9 | 52 | 0.00318 | 5 | -2.97521 |
| 1 | 0.25043 | 1 | 109 | 9 | 51 | 0.00028 | 3 | -3.30532 |
| 0 | 0.24514 | 0 | 113 | 9 | 54 | 0.00809 | 5 | -1.61377 |
| 1 | 0.49093 | 0 | 353 | 9 | 86 | 0.00482 | 6 | -1.63114 |
| 0 | 0.2008 | 0 | 679 | 8 | 69 | 0.0056 | 4 | -3.20775 |
| 0 | 0.53926 | 1 | 104 | 9 | 49 | 0.00371 | 3 | -2.93435 |
| 1 | 0.38069 | 1 | 108 | 9 | 53 | 0.00998 | 3 | -1.06223 |
| 0 | 0.45829 | 0 | 104 | 10 | 52 | 0.00087 | 2 | -3.34138 |
| 0 | 0.55563 | 0 | 109 | 10 | 52 | 0.00099 | 2 | -3.32156 |
| 0 | 0.3131 | 1 | 107 | 8 | 52 | 0.0011 | 2 | -3.22894 |
| 1 | 0.37403 | 1 | 108 | 9 | 53 | 0.00877 | 3 | -2.10981 |
| 1 | 0.21769 | 1 | 106 | 10 | 53 | 0.00696 | 2 | -2.0926 |
| 0 | 0.14983 | 1 | 105 | 8 | 51 | 0.00062 | 4 | -3.42681 |
| 1 | 0.18966 | 0 | 108 | 9 | 54 | 0.00941 | 3 | -1.58571 |
| 0 | 0.5363 | 1 | 681 | 9 | 69 | 0.00365 | 4 | -3.36806 |
| 0 | 0.32927 | 1 | 114 | 10 | 54 | 0.00705 | 5 | -2.16467 |
| 1 | 0.485 | 1 | 351 | 9 | 85 | 0.00919 | 5 | -2.16835 |
| 1 | 0.36979 | 0 | 116 | 9 | 53 | 0.00472 | 3 | -2.27359 |
| 0 | 0.32913 | 1 | 105 | 10 | 52 | 0.00546 | 2 | -1.9901 |
| 1 | 0.56244 | 0 | 135 | 8 | 83 | 0.00599 | 4 | -1.60558 |
| 0 | 0.32667 | 1 | 107 | 9 | 54 | 0.00466 | 2 | -2.48184 |
| 1 | 0.40397 | 0 | 135 | 7 | 85 | 0.00805 | 2 | -1.05282 |
| 0 | 0.50798 | 0 | 105 | 10 | 50 | 0.00585 | 2 | -1.41318 |
| 1 | 0.5767 | 1 | 112 | 9 | 53 | 0.00081 | 3 | -3.3231 |
| 1 | 0.34536 | 1 | 106 | 10 | 53 | 0.00642 | 3 | -1.70746 |
| 0 | 0.35831 | 0 | 136 | 7 | 83 | 0.00108 | 4 | -3.63163 |
| 0 | 0.40121 | 1 | 107 | 9 | 53 | 0.00702 | 5 | -2.08401 |
| 0 | 0.16599 | 1 | 353 | 10 | 84 | 0.00829 | 4 | -2.1307 |
| 0 | 0.15096 | 1 | 107 | 10 | 52 | 0.00954 | 2 | -1.05146 |
| 1 | 0.26299 | 0 | 135 | 7 | 85 | 0.00944 | 3 | -1.05272 |
| 1 | 0.22333 | 1 | 108 | 9 | 52 | 0.00278 | 3 | -3.01575 |
| 0 | 0.29898 | 0 | 136 | 10 | 84 | 0.00934 | 3 | -1.59114 |
| 1 | 0.38758 | 0 | 134 | 10 | 84 | 0.00941 | 3 | -1.60737 |
| 1 | 0.46434 | 1 | 134 | 5 | 85 | 0.00197 | 3 | -3.2531 |
| 0 | 0.31317 | 0 | 107 | 9 | 51 | 0.00117 | 2 | -3.29504 |
| 1 | 0.48507 | 0 | 114 | 9 | 53 | 0.00515 | 4 | -2.29615 |
| 1 | 0.50816 | 1 | 108 | 8 | 55 | 0.00695 | 4 | -2.17835 |
| 0 | 0.31236 | 0 | 114 | 9 | 54 | 0.00416 | 2 | -2.18348 |